\documentclass[
superscriptaddress,
showpacs,
nofootinbib,
amsmath,amssymb,
aps,
pra,
floatfix,longbibliography
]{revtex4-2}
\usepackage[T1]{fontenc}
\usepackage{graphicx}
\usepackage{dcolumn}
\usepackage[hypertexnames, breaklinks=true, bookmarksnumbered=true, bookmarksopen=true, colorlinks=true, linktocpage=true, citecolor=blue, urlcolor=magenta, linkcolor=magenta]{hyperref}
\usepackage{amsfonts,amsthm}
\usepackage{mathrsfs}
\usepackage{bbm} 
\usepackage{bm} 
\usepackage{mathtools} 
\usepackage[stretch=10]{microtype} 
\usepackage[normalem]{ulem} 
\usepackage{empheq}
\usepackage{color}
\usepackage{tcolorbox}
\usepackage{natbib}
\usepackage{pxfonts,txfonts}
\usepackage{tgpagella}

\allowdisplaybreaks

\allowdisplaybreaks

\begin{document}


\title{A study of the optimization problem on the combination of sectionalizing switches in power grid with quantum annealing}
\author{Masaya Takahashi}
\email{takahashi-mas@chodai.co.jp}
\author{Hiroaki Nishioka}
\author{Masahiro Hirai}
\author{Hidetaka Takano}
\affiliation{Chodai CO., LTD., Tokyo, Japan}

\begin{abstract}From the perspective of global warming, efficiency improvement of power grids is a pressing issue. Power grids have many switching devices to control the flow of electricity. Since there is a slight resistance in the wires and power consumption is proportional to the square of the current, the value of power loss on the wires changes depending on the combination of switch values that change the supply path of the current. The total number of switch combinations increases exponentially with the number of switches, and various algorithms have been studied to find the optimal combination of switch values. We propose a method to capture the switch combination problem in power grids as quadratic unconstrained binary optimization (QUBO) and derive an evaluation function to solve it using quantum annealing. The result is registered as a patent P6736787 at Japanese patent office. 
\end{abstract}

\maketitle

\section{Introduction \label{Sec:Introduction}}
In order to enhance the reliability, power grids are typically designed to allow for several configuration of distribution paths to increase redundancy. This is realized by manipulating on/off on a number of switches installed inside the power grid. The impedance of distribution line is typically quite small. However, there is still unignorable amount of power loss on it turning into heat. 
Therefore, a number of studies have been made to develop algorithms that reduce the computational complexity of problem\cite{Inoue2014,Gautam2020,Hayashi-2006a,Hayashi-2006b}.

Recently, quantum annealing has been paid attention as a solution to such optimization problems.
Quantum annealing is a type of computer that specializes in solving optimization problems\cite{Kadowaki1998,das2005quantum}. In recent years, as the number of qubits in quantum annealing machines has increased, there have been numerous attempts to solve real-world problems in various fields, such as transportation\cite{Volkswagen}. However, there has been no application on the optimization problem of switch configuration in the power grid yet.


In this paper, we formulate the evaluation function of which will be the input for quantum annealing machine to solve the problem. It is composed by the energy loss function on the distribution and the penalty functions in terms of values of switches as binary valuables.

\begin{figure}
    \centering
    \includegraphics[scale=0.7]{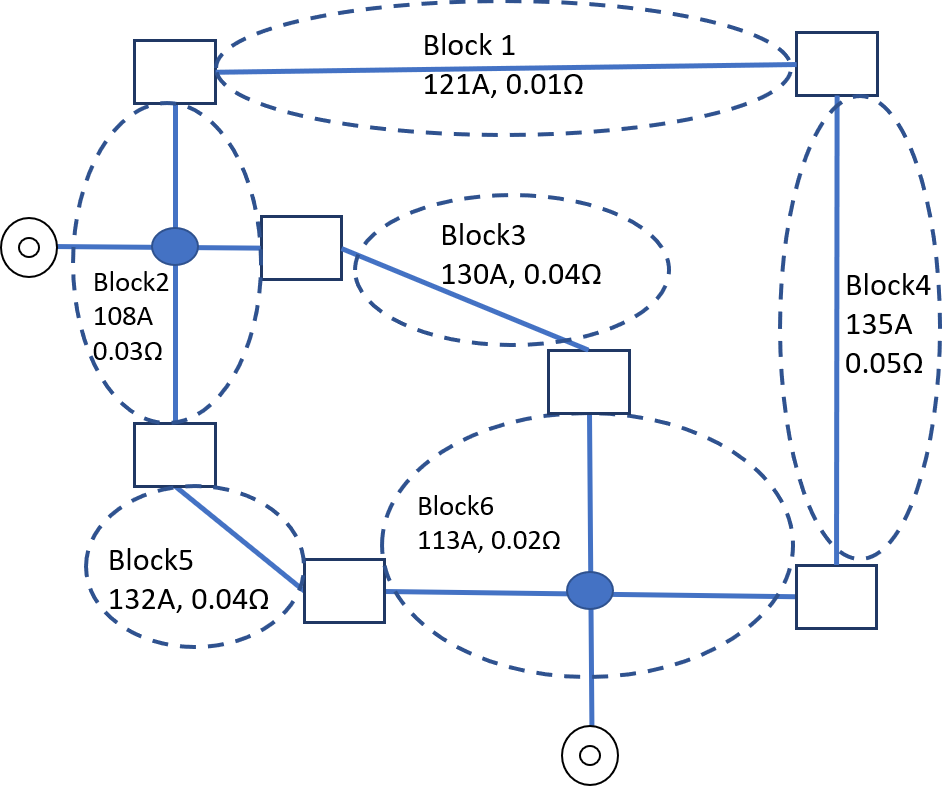}
    \caption{Double circle and square represent feeder and switch respectively. Switches divide power grid into blocks represented by dotted circle which has load current and resistance.}
    \label{fig:powergrid}
\end{figure}

A power grid may have multiple feeders and switches(see Fig.~\ref{fig:powergrid}). When a switch is open, electric current does not go through it. Vice versa, electric current go through a switch if it is closed. 
A power grid can be partitioned into blocks by switches. After numbering blocks, a switch is uniquely determined by two adjacent blocks. Let $q_{ij}$ represents the open/closed state of switch between block $i$ and block $j$. Namely, $q_{ij} = 0$ indicates the switch between block $i$ and $j$ is open which means that the electric current cannot go through the switch. Meanwhile $q_{ij} = 1$ indicates the switch is closed which means that the electric current can go through the switch. We denote the resistance of block $i$ as $R_i$ and its load current as $I_i$. 
Let $S$ be the total number of blocks. To make $q_{ij}$ well defined for all combination of $i, j \in S$, let $q_{ij} = 0$ if there is no switch between blocks. That is, we consider there is an imaginary switch always open between them(electric current never flow). 
Also, let $q_{ii}$ represents whether the block $i$ is connected to feeder of voltage $V_i$. Then, $q_{ij}$ is well defined for all combination of $i,j \in S$. 

\section{Formulation}
Now, let us suppose that the switch between block 1 and 2 in Fig.~\ref{fig:powergrid} is closed. Block 2 supplies electricity to block 1, so the total current flowing through block 2 is increased by the load current of block 1. Moreover, if the two switches between 2, 1 and 4 are closed, the total current flowing through 2 is increased by the load current of block 1 and 4 as block 2 supplies electricity to both block 1 and 4. With the above considerations, the total current $I'_i$ flowing through block $i$ can be represented as follows.

\begin{align}\label{eq:current}
I'_i = I_i + \sum_{j \not = i}q_{ij} \prod_{k \not = i} \overline{q_{jk}}I_j + \sum_{\substack{ l\not = i \\ m \not  = i,l}}q_{il} q_{lm} \overline{q_{ll}} ~  \overline{q_{mm}}(I_l+I_m) 
\end{align}
where $\overline{q_{ij}} = 1 -q_{ij}$. $I'_i$ is divided into three parts. The first term represents the load current $I_i$ of block $i$ itself.

The second term is the load current to be supplied to adjacent block which is not connected to other blocks than block $i$. The factor $q_{ij}$ in the second term becomes $1$ only when block $j$ is connected to block $i$. The factor $\prod_{k\not = i} \overline{q_{jk}}$ in the second term becomes $1$ only when block $j$ is not connected to any other block $k$ except for $i$, and the power feeder is not connected. Therefore, the second term as a whole adds up the load currents of adjacent blocks that satisfy the above conditions.

The third term is the the load current to be supplied to adjacent two-subsequent blocks which does not connected to other blocks than block $i$. The factor $q_{il}$ in the third term becomes $1$ only when block $l$ is connected to block $i$. Similarly, $q_{lm}$ becomes $1$ only when block $m$ is connected to block $l$. Therefore, the product $q_{il}q_{lm}$ checks whether two blocks are connected in a row starting from block $i$. The factor $\overline{q_{ll}}\overline{q_{mm}}$ becomes $1$ only when the intermediate block $l$ and the end block $m$ are not connected to the feeder. Since the maximum connection condition which will be described later, prevent to pick up the configuration where more 4 blocks are connected, we do not check if block $l$ and $m$ are connected other blocks than block $i$, which helps us to simplifies the equation. The energy loss function is given by following.
\begin{align}\label{eq:energyloss}
P_{\rm power} = \sum_i R_i (I'_i)^2   
\end{align}
where $I'_i$ is total current of $i$th block.

Due to physical constraint, the power grid must satisfy conditions, such as radial condition, maximum connection condition (up to three blocks), no blackout condition, maximum current condition, maximum voltage condition, and minimum voltage condition. For each of these conditions, we introduce a corresponding penalty function. When a condition is satisfied, the corresponding penalty function is defined to be zero. When a condition is not satisfied, the penalty function should take a large number compared to the energy loss function, so that the evaluation function also becomes large regardless the energy loss. 
Since such a large number of the evaluation function represents 
a state corresponding high energy in the quantum annealing machine, the probability of such states being selected by the annealing solver becomes very small. Therefore we would obtain a configuration where conditions are satisfied. 

The radial condition is a condition in which blocks connected to the feeder must not be electrified with each other. The radial condition function $P_{\rm radial}$ is represented as follows:
\begin{align}
P_{\rm radial} = \sum_{i}\sum_{j\not = i} q_{ii}q_{jj}[ q_{ij} + \sum_k q_{ik} q_{kj}]C_{\rm penalty}
\end{align}
where $C_{\rm penalty}$ is a penalty constant which should be a suitable large number. The factor $q_{ii}q_{jj}$ becomes 1 only when both block $i$ and block $j$ are connected to their feeders. The expression in the bracket becomes 1 when block $i$ and block $j$ are connected adjacently or through an intermediate block. Therefore, this whole function becomes a representation of the radial condition.

The maximum connection condition (3 blocks) is a condition that states that no more than 4 blocks can be connected(powered) at once. Even if three blocks A, B, and C are connected in a circular manner, we consider them to be connected by four or more blocks (i.e., A B C A ...) and exclude them from the solution candidates based on this condition. Because of this condition, it is not necessary to consider connections of four or more blocks in a loop. The maximum connection condition function is expressed as follows:

\begin{align}
P_{\rm max conn.} = \left[\sum_{i}\sum_{j \not = i}\sum_{\substack{k \not =j,i}}\sum_{\substack{l \not = j,k \\ l \geq i}} q_{ij} q_{jk} q_{kl}(1-q_{ii}q_{kk})(1-q_{jj}q_{ll})  + \sum_{i}\sum_{j \not = i}\sum_{\substack{k \not =j \\ k > i}} \sum_{\substack{l \not =j \\ l > i \\ l > k}}q_{ij} q_{jk} q_{jl}(1-q_{ii}q_{kk})(1-q_{kk}q_{ll})(1-q_{ll}q_{ii}) \right]C_{\rm penalty}
\end{align}

The first term corresponds to the case where four blocks are connected in a straight line. $q_{ij} q_{jk} q_{kl}$ will be $1$ only if block $i,j,k,l$ are connected. Notice $l$ can be the same number with $i$ to avoid a loop connection. Also, we assume $l$ is greater than or equal to $i$ because the fourth block number can be counted in reverse order if it is smaller than the first block number. In addition, when both the first block $i$ and the third block $k$ are connected to a feeder, they can be excluded from the maximum connection condition based on the aforementioned radial condition. Similarly, if both the second block $j$ and the fourth block $l$ are connected to a feeder, they are also excluded. These correspond to $(1-q_{ii}q_{kk})$ and $(1-q_{jj}q_{ll})$, respectively. Here, it is assumed that the powered blocks are separated by at least one block.

The second term corresponds to the case where four blocks are connected in a T-shape. Since the three blocks at the end of the T-shape are symmetrically positioned, we count only those with $k$ greater than $j$ and $l$ greater than both $k$ and $j$ to prevent double counting. Cases where two of the three blocks at the end of the T-shape are connected to a feeder are excluded because they overlap with the radiation condition.

The no blackout condition is a condition that all blocks must be connected to at least one feeder. It is represented by the following equation:
\begin{align}
P_{\rm blackout} = \sum_i \overline{q_{ii}} \prod_{j \not = i}(1 - q_{jj} q_{ij}) \prod_{\substack{k \not = i \\ l \not = i,k}}(1 - q_{ll} q_{ik} q_{kl}) C_{\rm penalty}
\end{align}
The factor $\overline{q_{ii}}$ is equal to 0 when a feeder is connected to block $i$. The factor $(1 - q_{jj}q_{ij})$ is equal to 0 when block $i$ and block $j$ are connected and block $j$ is connected to a feeder. If there exists at least one such block$ j$, the product is 0, causing the entire equation to be equal to 0 for block $i$. The factor $(1 - q_{ll} q_{ik} q_{kl})$ is equal to 0 when block $i,k$, and $l$ are connected (where block $l$ is the end terminal) and block $l$ is connected to a feeder.

The maximum current condition states that the current through in each block must be smaller than the certain amount due to the line's capacity for electric current. Let $I_i^{\rm max}$ denote the maximum allowable current for block $i$, then $I'_i$ must satisfy $I'_i < I_i^{\rm max}$. This condition is expressed as follows, where $L$ is an appropriate integer:
\begin{align}
P_{\rm current} &=\sum_i \bigg (\frac{ I'_i}{I_i^{\rm max}} \bigg )^L C_{\rm{penalty}}
\end{align}
If $I'_i < I_i^{\rm max}$, then setting $L$ to enough large causes $\bigg (\frac{ I'_i}{I_i^{\rm max}} \bigg )^L$ to approach zero, whereas if $I'_i \geq I_i^{\rm max}$, then $\bigg (\frac{ I'_i}{I_i^{\rm max}} \bigg )^L$ becomes  large. 

When current flows through a block, voltage drop occurs according to Ohm's law. However, the voltage that reaches each block must be within a certain range. We achieve this by two conditions: the maximum voltage and the minimum voltage. Both of these conditions must be satisfied. To formalize these conditions, several quantities need to be defined. First, we define the voltage drop $V^{\rm drop}_i$ occurred in block $i$  by $V^{\rm drop}_i = I'_i R_i$. The voltage of the feeder powering block $i$ is $F_i$, and the voltage at block $i$ is denoted by $V_i$. It can be expressed as follows:

\begin{align}
V_i = q_{ii}F_i + \overline{q_{ii}}\left[\sum_{j \not = i} q_{ij}q_{jj}(F_j - V^{\rm drop}_j) + \sum_{\substack{k \not = i \\ l \not = k}} q_{ik}q_{kl}q_{ll}(F_l - V^{\rm drop}_k - V^{\rm drop}_l)\right]
\end{align}

Let us denote the maximum voltage allowable for block $i$ by $V^{\rm max}_i$. Then, the maximum voltage condition $P_{\rm max ~ V}$ is expressed as follows using the same structure as the maximum current condition:

\begin{align}\label{uppervol}
P_{\rm max ~ V} = \sum_i \bigg (\frac{V_i}{V^{\rm max}_i} \bigg )^L C_{\rm penalty}
\end{align}

Next, we formulate the minimum voltage condition. Here, we want to create a conditional expression similar to the form of Equation \ref{uppervol}. However, if a fraction has a variable $q$ in its denominator, the form of function is no longer QUBO unless we convert them. It is not known if there is an easy way to convert such fraction to QUBO.  Therefore, we will create the conditional expression by comparing the amount of voltage drop from the reference voltage where we do not have variables in the denominator. First, let the reference voltage $F_{\rm max}$ be a value such that $F_{\rm max} > F_i$ for any $i$. Also, let $\overline{V}^{\rm drop}_i$ be the cumulative voltage drop from the reference voltage $F_{\rm max}$ at block $i$ from the feeder which supplies electricity to block $i$. Then,  $\overline{V}^{\rm drop}_i$ becomes as follows:
\begin{align}
\overline{V}^{\rm drop}_i = q_{ii}(F_{\rm max} - F_i)  + \overline{q_{ii}}\left[\sum_{j \not = i} q_{ij}q_{jj}((F_{\rm max} - F_j) + V^{\rm drop}_j) + \sum_{\substack{k \not = i \\ l \not = k}} q_{ik}q_{kl}q_{ll}( (F_{\rm max} - F_l) + V^{\rm drop}_k + V^{\rm drop}_l)\right ]
\end{align}
Next, let $\overline{V}^{\rm drop \ max}_i$ be the maximum allowable cumulative voltage drop at block $i$. Then, we must have $\overline{V}^{\rm drop \ max}_i > \overline{V}^{\rm drop}_i$. Therefore, the minimum voltage condition $P_{\rm min \ V}$ can be expressed as follows:
\begin{align}
P_{\rm min \ V} = \sum_i \bigg (\frac{\overline{V}^{\rm drop}_i}{\overline{V}^{\rm drop \ max}_i} \bigg )^L C_{\rm{penalty}}
\end{align}

Putting these condition functions and the energy loss function together, the evaluation function $P_{\rm total}$  becomes as follows.
\begin{align}
P_{\rm total} = P_{\rm power} + P_{\rm radial} + P_{\rm maxconn.} + P_{\rm blackout} + P_{\rm current} + P_{\rm max V} + P_{\rm min V} 
\end{align}

\section{example}
\begin{figure}[h]
\centering
\includegraphics[scale=0.7]{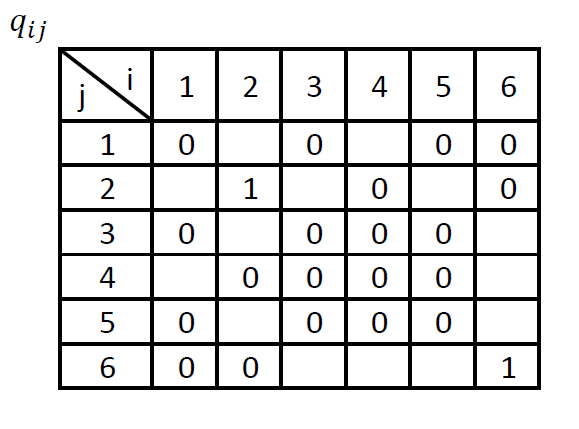}
\caption{Most of elements in this table is constant. $0$ represents the switch is open for non-diagonal element or a feeder is not connected for diagonal elements. $1$ represents the switch is closed for non-diagonal element or a feeder is connected for diagonal elements.}
\label{fig:table}
\end{figure}

In this section, we demonstrate the calculation of the total current and some constraint conditions in the example of Figure 1. First, the presence or absence of switches and feeders does not change. Therefore, some variables can be treated as constants in essence. For example, since no feeder is connected to block 1, $q_{11} = 0$ always holds (See Fig~\ref{fig:table}). 

The total currents are given by the following equations:
\begin{align}
I'_1 =&I_1 + q_{14}\overline{q_{46}}I_4\\
(I'_1)^2 =&I_1^2 +2q_{14}\overline{q_{46}}I_4I_1 + q_{14}\overline{q_{46}}I_4^2\\
I'_2 =&I_2 + q_{21}\overline{q_{14}}I_1 + q_{23}\overline{q_{36}}I_3 + q_{25}\overline{q_{56}}I_5 + q_{ 21}q_{14}(I_1 + I_4)\\
(I'_2)^2 =&I_2^2 + q_{21}\overline{q_{14}}I_1^2 + q_{23}\overline{q_{36}}I_3^2+ q_{25}\overline {q_{56}}I_5^2 + q_{21}q_{14}(I_1 + I_4)^2 \\
+&2[q_{21}\overline{q_{14}}I_2I_1 + q_{23}\overline{q_{36}}I_2I_3 + q_{25}\overline{q_{56}}I_2I_5 + q_{21}q_ {14}I_2(I_1 + I_4)\\
+& q_{21}\overline{q_{14}}I_1q_{23}\overline{q_{36}}I_3 + q_{21}\overline{q_{14}}I_1q_{25}\overline{q_{56 }}I_5+q_{23}\overline{q_{36}}I_3q_{25}\overline{q_{56}}I_5\\
+& q_{23}\overline{q_{36}}I_3q_{21}q_{14}(I_1 + I_4) + q_{25}\overline{q_{56}}I_5q_{21}q_{14}(I_1 + I_4) ]\\
I'_3 =& I_3\\
(I'_3)^2 =& I_3^2 \\
I'_4 =& I_4 + q_{41}\overline{q_{12}}I_1\\
(I'_4)^2 =& I_4^2 + 2q_{41}\overline{q_{12}}I_1I_4+q_{41}\overline{q_{12}}I_1^2\\
I'_5 =& I_5\\
(I'_5)^2 =& I_5^2\\
I'_6 =&I_6 + q_{63}\overline{q_{32}}I_3 + q_{64}\overline{q_{41}}I_4 + q_{65}\overline{q_{52}}I_5 + q_{ 64}q_{41}(I_4 + I_1)\\
(I'_6)^2 =&I_6^2 + q_{63}\overline{q_{32}}I_3^2 + q_{64}\overline{q_{41}}I_4^2+ q_{65}\overline {q_{52}}I_5^2 + q_{64}q_{41}(I_4 + I_1)^2 \\
+&2[q_{63}\overline{q_{32}}I_6I_3 + q_{64}\overline{q_{41}}I_6I_4 + q_{65}\overline{q_{52}}I_6I_5 + q_{64}q_ {41}I_6(I_1 + I_4)\\
+& q_{63}\overline{q_{32}}I_3q_{64}\overline{q_{41}}I_4 + q_{63}\overline{q_{32}}I_3q_{65}\overline{q_{52 }}I_5+q_{64}\overline{q_{41}}I_4q_{65}\overline{q_{52}}I_5\\
+& q_{63}\overline{q_{32}}I_3q_{64}q_{41}(I_4 + I_1) + q_{65}\overline{q_{52}}I_5q_{64}q_{41}(I_4 + I_1) ]
\end{align}
We used $q_{ij}\overline{q_{ij}} = 0$ to simplifies the results. Radial conditions are given below.
\begin{align}
P_{\rm radial} = [q_{23}q_{36} + q_{25}q_{56}]C_{\rm penalty}
\end{align}
The maximum connection condition is given below.
\begin{align}
P_{\rm connection} = [q_{14}q_{46}q_{63} +q_{14}q_{46}q_{65} +q_{32}q_{21}q_{14} +q_{41 }q_{12}q_{25} +q_{12}q_{23}q_{25} +q_{36}q_{64}q_{65} ] C_{\rm penalty}
\end{align}
The uninterruptible conditions are given by
\begin{align}
P_{\rm blackout} = [\overline{q_{12}}(1 -q_{14}q_{46}) + \overline{q_{32}}\overline{q_{36}} + \overline{q_ {46}}(1 -q_{41}q_{12})+ \overline{q_{52}}\overline{q_{56}}]\rm{C_{penalty}}.
\end{align}

In addition, we attach a Python code that evaluates the evaluation function for this sample distribution network as supplementary material. Sending the computed evaluation function to the annealing machine, the reader may obtain the configuration (q12, q14, q23, q25, q36, q46, q56) = (0, 1, 0, 0, 1, 1, 1) as the solution. The input data needs to be written directly input into the code for simplicity. By modifying the input, the reader can obtain the optimal combination of switches for other distribution network.

\bibliography{power}

\end{document}